\documentclass[referee]{aa} 
\usepackage{natbib}
\bibpunct{(}{)}{;}{a}{}{,} 

\usepackage{longtable} 
\usepackage{graphicx}
\usepackage[varg]{txfonts}
\begin{document}
   \title{Optical observations of NEA 3200 Phaethon (1983 TB) during the 2017 apparition\thanks{Photometric data will be available in electronic form at the CDS via anonymous ftp to cdsarc.u-strasbg.fr (130.79.128.5) or via http://cdsarc.u-strasbg.fr/viz-bin/qcat?J/A+A/volume/page}}

   \author{M.-J. Kim \inst{1} \and
	  H.-J. Lee \inst{1,2} \and
	  S.-M. Lee \inst{1,2} \and
	  D.-H. Kim \inst{1,2} \and
	  F. Yoshida \inst{3} \and
	  P. Bartczak \inst{4} \and
	  G. Dudzi\'{n}ski \inst{4} \and
      J. Park \inst{1} \and
      Y.-J. Choi \inst{1,5} \and
      H.-K. Moon \inst{1} \and
      H.-S. Yim \inst{1} \and
      J. Choi \inst{1,5} \and
      E.-J. Choi \inst{1} \and
      J.-N. Yoon \inst{6} \and
      A. Serebryanskiy \inst{7} \and
      M. Krugov \inst{7} \and
      I. Reva \inst{7} \and
      K. E. Ergashev \inst{8} \and
      O. Burkhonov \inst{8} \and
      S. A. Ehgamberdiev \inst{8} \and
      Y. Turayev \inst{8} \and
      Z.-Y. Lin \inst{9} \and
      T. Arai \inst{3} \and
      K. Ohtsuka \inst{10} \and
      T. Ito \inst{11} \and
      S. Urakawa \inst{12} \and
      M. Ishiguro \inst{13}
      }

   \institute{Korea Astronomy and Space Science Institute, 776, Daedeokdae-ro, Yuseong-gu, Daejeon 34055, Korea \\
              \email{skarma@kasi.re.kr}\label{inst1} \and
              Chungbuk National University, 1 Chungdae-ro, Seowon-gu, Cheongju, Chungbuk 28644, Korea \and
              Planetary Exploration Research Center, Chiba Institute of Technology, 2-17-1 Tsudanuma,Narashino, Chiba 275-0016, Japan \and
              Astronomical Observatory Institute, Faculty of Physics, Adam Mickiewicz University, S{\l}oneczna 36, 60-286 Pozna{\'n}, Poland \and
              University of Science and Technology, 217, Gajeong-ro, Yuseong-gu, Daejeon 34113, Korea \and
              Chungbuk National University Observatory, 802-3 Euntan-ri, Jincheon-gun, Chungcheongbuk-do, Korea \and
              Fesenkov Astrophysical Institute, Observatory 23, 050020 Almaty, Kazakhstan \and
              Ulugh Beg Astronomical Institute of the Uzbekistan Academy of Sciences, 33 Astronomicheskaya str., Tashkent, 100052, Uzbekistan \and
              Institute of Astronomy, National Central University, No. 300, Zhongda Rd., Zhongli Dist., Taoyuan City 32001, Taiwan \and
              Tokyo Meteor Network, Daisawa 1–27–5, Setagaya-ku, Tokyo 155–0032, Japan \and
              National Astronomical Observatory of Japan, Osawa 2-21-1, Mitaka, Tokyo 181-8588, Japan \and
              Japan Spaceguard Association, Bisei Spaceguard Center, 1716-3 Okura, Bisei-cho, Ibara, Okayama 714-1411, Japan \and
              Seoul National University, 1 Gwanak-ro, Gwanak-gu, Seoul 08826, Korea
              }

   \date{Received June 08, 2018; accepted  Aug 30, 2018}

 
  \abstract
   {The near-Earth asteroid 3200 Phaethon (1983 TB) is an attractive object not only from a scientific viewpoint but also because of JAXA's DESTINY$^{+}$\thanks{Demonstration and Experiment of Space Technology for INterplanetary voYage, Phaethon fLyby and dUSt science} target. The rotational lightcurve and spin properties were investigated based on the data obtained in the ground-based observation campaign of Phaethon.}
   {We aim to refine the lightcurves and shape model of Phaethon using all available lightcurve datasets obtained via optical observation, as well as our time-series observation data from the 2017 apparition.}
   {Using eight 1–2-m telescopes and an optical imager, we acquired the optical lightcurves and derived the spin parameters of Phaethon. We applied the lightcurve inversion method and SAGE (Shaping Asteroids with Genetic Evolution) algorithm to deduce the convex and non-convex shape model and pole orientations.}
   {We analysed the optical lightcurve of Phaethon and derived a synodic and a sidereal rotational periods of 3.6039 h, with an axis ratio of $a/b = 1.07$. The ecliptic longitude ($\lambda_\mathrm{p}$) and latitude ($\beta_\mathrm{p}$) of the pole orientation were determined as (308, -52) and (322, -40) via two independent methods. A non-convex model from the SAGE method, which exhibits a concavity feature, is also presented.}
   {}

   \titlerunning{Phaethon}
   \authorrunning{Myung-Jin Kim et al}

   \maketitle
%

\section{Introduction}
The near-Earth asteroid (NEA) (3200) Phaethon (1983 TB) (hereinafter referred to as Phaethon) is the target of the DESTINY$^{+}$ mission, which is an Epsilon-class programme, and is currently under Phase-A study by JAXA (Japan Aerospace Exploration Agency)/ISAS (Institute of Space and Astronautical Science). Phaethon is classified as a member of the Apollo asteroidal group with a semi-major axis greater than that of the Earth. In addition, it is called as Mercury-crosser asteroid with the small perihelion distance of only 0.14 AU. It is also categorised as a potentially hazardous asteroid; the Earth minimum orbit intersection distance is 0.01945 AU. The spectral type of Phaethon is known as B-type \citep{green,binzel2001,binzel2004,bus}, which is a sub-group of C-complex that is attributed to primitive volatile-rich remnants from early solar system. The asteroid (24) Themis - a typical B-type asteroid - was recently discovered to have H$_\mathrm{2}$O ice and organic matter on its surface \citep{rivkin,campins}. Phaethon is thus one of the most remarkable NEAs, not only because of its spectral type but also because of its extraordinary connection with the Geminids meteor shower that occurs every mid-December \citep[][and references therein]{gustafson,williams,jenniskens}. 

\begin{table*}
\caption{Observatory and instrument details.}
\label{table:1}
\centering
\begin{tabular}{c c c c c c c}     
\hline\hline
Telescope\tablefootmark{a} & $\lambda$\tablefootmark{b} & $\phi$\tablefootmark{b} & Altitude & Instrument\tablefootmark{c} & Pixel scale & Observer\tablefootmark{d} \\
     &    &    & [m] & [CCD] & [$''$pix$^{-1}$] & \\
\hline
 SLT 0.4 m & 120:52:25 & +23:28:07 & 2,879.0 & e2v CCD42-40 & 0.79 & ZYL \\
 OWL 0.5 m & 249:12:38 & +32:26:32 & 2,769.5 & FLI 16803 & 0.98 & JC, EJC \\
 SOAO 0.6m	  & 128:27:27	& +36:56:04	& 1,354.4 & 	FLI 16803	     & 0.45	 &  TSJ, GYH \\      
CBNUO 0.6 m	& 127:28:31	& +36:46:53	& 87.0    &	STX-16803	       & 1.05	 &  SML, JNY \\      
MAO 0.6 m	  & 66:53:44	& +38:40:24	& 2,578.2 & 	FLI IMG1001E	 & 0.68	 &  EK, OB, SAE, YT \\
TShAO 1.0 m	& 76:58:18	& +43:03:26	& 2,723.5 &	Apogee Alta F16M	& 0.56 &	AS, MK, IR \\     
LOAO 1.0 m	& 249:12:41	& +32:26:32	& 2,776.0 &	e2v 4K CCD	     &  0.80 &	FY, JHY, IKB  \\ 
BOAO 1.8 m	& 128:58:36	& +36:09:53	& 1,143.0 & 	e2v 4K CCD	   &  0.43 &	MJK, JP   \\     
\hline
\end{tabular}
\tablefoot{
\tablefoottext{a}{Abbreviations: SLT = Lulin Super Light Telescope, OWL = Optical Wide-field patroL, SOAO = Sobaeksan Optical Astronomy Observatory, CBNUO = ChungBuk National University Observatory, MAO = Maidanak Astronomical Observatory, TShAO = Tian Shan Astronomical Observatory, LOAO = Lemonsan Optical Astronomy Observatory, BOAO = Bohyunsan Optical Astronomy Observatory} 
\tablefoottext{b}{Eastern longitude and geocentric latitude of each observatory} 
\tablefoottext{c}{FLI 16803 in SOAO, e2v 4K CCD and SI 4K CCD were configured with 2 $\times$ 2 binning} 
\tablefoottext{d}{Observer: ZYL = Zhong-Yi Lin, JC = Jin Choi, EJC = Eun-Jung Choi, TSJ = Taek-Soo Jung, GYH = Gi-Young Han, SML = Sang-Min Lee, JNY = Joh-Na Yoon, EK = Ergashev Kamoliddin, OB = Otabek Burkhonov, SAE = Shuhrat A. Ehgamberdiev, YT = Yunus Turayev, AS = Alexander Serebryanskiy, MK = Maxim Krugov, IR = Inna Reva, FY = Fumi Yoshida, JHY = Jae-Hyuk Yoon, IKB = In-Kyung Baek, MJK = Myung-Jin Kim, JP = Jintae Park}
}
\end{table*}

For this reason, various investigations of the physical properties of Phaethon have been conducted. Regarding the rotational properties, the most recent results of lightcurve observations indicate that Phaethon has a rotational period of 3.604 h \citep{wisniewski,pravec,krugly,ansdell,warner2015,schmidt}. On the basis of the lightcurve analysis, Phaethon is regarded as having a nearly spherical shape with a small lightcurve amplitude of 0.1 – 0.2. \citet{ansdell} derived $\lambda=85\degr \pm 13\degr$ and $\beta=-20\degr \pm 10\degr$, while \citet{hanus2016} obtained a convex shape model of Phaethon and the pole axis of $\lambda=319\degr \pm 5\degr$ and $\beta=-39\degr \pm 5\degr$ according to previous and newly obtained lightcurves, where $\lambda$ and $\beta$ are the ecliptic longitude and latitude of the pole orientation, respectively. 

The observation window for Phaethon at the end of 2017 was a good opportunity to acquire high-quality dense photometric data, as the asteroid passed by the Earth at a lunar distance (LD) of only 27 LD on 16 December 2017, which was the closest approach in 40 years. We performed a photometric observation campaign for Phaethon between the Asian and American continents during the 2017 apparition to investigate its rotational properties and refine the pole solution. In this paper, we outline our optical observations, data reduction, and analysis. We derived the rotational period and peak-to-peak variation from the lightcurve. Furthermore, we deduced the pole orientation and shape model with not only a convex model based on the lightcurve-inversion method \citep{kaasalainen2001a,kaasalainen2001b} but also a non-convex model using the SAGE (Shaping Asteroids with Genetic Evolution) algorithm \citep{bartczak}. 
\begin{table*}
\caption{Observational circumstances.}
\label{table:2}
\centering
\begin{tabular}{c c c c c c c c c c c c}     
\hline\hline
UT date & RA & DEC & L$_\mathrm{PAB}$ & B$_\mathrm{PAB}$ & $\alpha$ & r & $\Delta$ & V & Telescope & Seeing & Sky \\
(DD/MM/YY) & [hr] & [\degr] & $ [\degr] $ & [\degr] & [\degr] &  [AU]  & [AU] & [Mag] &   & [\arcsec] & condition \\
\hline
11.4/11/2017 & 106.44   & +35.25 & 87.1 & 9.5    & 33.1 & 1.496 & 0.695 & 16.08 & LOAO  & 3.2 & Cirrus \\
12.4/11/2017 & 106.49   & +35.35 & 87.2 & 9.5    & 32.9 & 1.485 & 0.675 & 15.99 & LOAO  & 2.7 & Cirrus \\
13.4/11/2017 & 106.52   & +35.45 & 87.3 & 9.6    & 32.8 & 1.473 & 0.654 & 15.90 & LOAO  & 2.7 & Cirrus \\
16.8/11/2017 & 106.47   & +35.86 & 87.6 & 9.6    & 32.1 & 1.432 & 0.583 & 15.57 & TShAO & 3.2 & Clear  \\
19.8/11/2017 & 106.22   & +36.29 & 87.8 & 9.6    & 31.3 & 1.396 & 0.522 & 15.26 & TShAO & 2.5 & Clear  \\
19.9/11/2017 & 106.22   & +36.30 & 87.8 & 9.6    & 31.3 & 1.395 & 0.522 & 15.25 & MAO     & 2.3 & Clear  \\
20.8/11/2017 & 106.08   & +36.45 & 87.9 & 9.6    & 31.1 & 1.384 & 0.503 & 15.15 & TShAO & 2.3 & Clear  \\
20.9/11/2017 & 106.07   & +36.46 & 87.9 & 9.6    & 31.0 & 1.383 & 0.502 & 15.14 & MAO     & 2.2 & Clear  \\
21.8/11/2017 & 105.91   & +36.62 & 87.9 & 9.7    & 30.7 & 1.371 & 0.483 & 15.03 & TShAO & 2.8 & Clear  \\
21.9/11/2017 & 105.90   & +36.63 & 87.9 & 9.7    & 30.7 & 1.370 & 0.482 & 15.02 & MAO     & 1.8 & Clear  \\
22.8/11/2017 & 105.70   & +36.81 & 87.9 & 9.7    & 30.4 & 1.359 & 0.463 & 14.91 & TShAO & 3.6 & Clear  \\
22.9/11/2017 & 105.69   & +36.82 & 87.9 & 9.7    & 30.4 & 1.358 & 0.462 & 14.90 & MAO     & 1.7 & Clear  \\
23.9/11/2017 & 105.44   & +37.02 & 88.0 & 9.7    & 30.0 & 1.345 & 0.442 & 14.78 & MAO     & 1.7 & Cirrus \\
24.7/11/2017 & 105.21   & +37.18 & 88.0 & 9.7    & 29.7 & 1.335 & 0.427 & 14.68 & SOAO  & 4.8 & Cirrus \\
26.7/11/2017 & 104.48   & +37.65 & 87.9 & 9.8    & 28.8 & 1.309 & 0.389 & 14.40 & SOAO  & 4.2 & Clear  \\
27.9/11/2017 & 103.91   & +37.98 & 87.8 & 9.9    & 28.2 & 1.293 & 0.365 & 14.22 & MAO     & 3.7 & Cirrus \\
01.6/12/2017 & 101.34   & +39.22 & 87.3 & 10.1 & 25.9 & 1.242 & 0.294 & 13.60 & BOAO    & 1.7 & Clear  \\
07.6/12/2017 & 91.75    & +42.39 & 84.4 & 11.0 & 20.7 & 1.156 & 0.185 & 12.28 & CBNUO   & 3.5 & Clear  \\
15.2/12/2017 & 33.6       & +40.54 & 64.1 & 14.2 & 41.4 & 1.040 & 0.076 & 10.72 & OWL     & 4.8 & Clear  \\
15.6/12/2017 & 26.86    & +38.13 & 61.1 & 14.3 & 46.6 & 1.033 & 0.073 & 10.76 & SLT       & 4.2 & Cirrus \\
16.2/12/2017 & 18.28    & +34.17 & 57.1 & 14.3 & 54.1 & 1.023 & 0.070 & 10.86 & OWL       & 4.3 & Clear  \\
17.1/12/2017 & 5.23       & +26.07 & 50.1 & 13.7 & 67.2 & 1.008 & 0.068 & 11.18 & OWL     & 4.5 & Clear  \\
\hline
\end{tabular}
\tablefoot{$UT$ date corresponding to the mid time of the observation, J2000 coordinates of Phaethon ($RA$ and $DEC$), Phase Angle Bisector (PAB) - the bisected arc between the Earth-asteroid and Sun-asteroid lines - ecliptic longitude ($L_\mathrm{PAB}$) and ecliptic latitude ($B_\mathrm{PAB}$), the solar phase angle ($\alpha$), the helicentric ($r$) and the topocentric distances ($\Delta$), the apparent predicted magnitude ($V$), average seeing and sky condition.}
\end{table*}

\section{Observations}
Photometric observations of Phaethon were conducted for a total of 22 nights with several 1–2-m-class telescopes equipped with CCD (Charge-Coupled Device) cameras. As the predicted apparent magnitude of the asteroid during the period between early November and mid-December 2017 was 11–16 magnitudes, the 1–2-m-class telescopes allowed us to obtain a lightcurve with a sufficient signal-to-noise (S/N) ratio. For the sake of securing the target visibility (that is, maintaining the declination coordinate of Phaethon at larger than 25 degrees and performing continuous observations), observatories in the Asian and American continents located in the northern hemisphere were used. We employed the Sobaeksan Optical Astronomy Observatory (SOAO) 0.6-m, ChungBuk National University Observatory (CBNUO) 0.6-m, and Bohyunsan Optical Astronomy Observatory (BOAO) 1.8-m telescopes in Korea; the Optical Wide-field patroL (OWL) 0.5-m and Lemonsan Optical Astronomy Observatory (LOAO) 1.0-m telescopes in Arizona, USA; the Tian Shan Astronomical Observatory (TShAO) 1.0-m telescope in Kazakhstan; the Maidanak Astronomical Observatory (MAO) 0.6-m North telescope in Uzbekistan; and the Lulin Super Light Telescope (SLT) 0.4-m telescope in Taiwan. The details of the observatories, including the instruments, are shown in Table 1. Out of 22 nights of observations, all images acquired using the LOAO 1.0 m telescope were obtained in the non-sidereal tracking mode corresponding to the predicted motion of the object, whereas the other telescopes were guided at sidereal rates. During the observations made in the sidereal rate, the maximum exposure time did not exceed 200 sec. The exposure time was determined by two factors. The apparent motion of the asteroid had to be less than the nightly average full width at half maximum (FWHM) of the stellar profiles at each observatory, and the S/N ratio of the object had to be >70.

Details of the observational circumstances are shown in Table 2. The phase angle bisector (PAB) is the bisected arc between the Earth-asteroid and Sun-asteroid lines that is expressed in ecliptic longitude ($L_\mathrm{PAB}$) and ecliptic latitude ($B_\mathrm{PAB}$). Observation conducted in a wide range of PABs is essential for deriving the pole orientation and three-dimensional (3D) shape model. For this purpose, we observed Phaethon at different geometries between the Earth and the asteroid with respect to the Sun. The viewing geometry - especially during the close approaching phase - dramatically changed around UT 23:00 on 16 December 2017. The observation using the OWL 0.5-m telescope in Arizona, USA was conducted before and after the closest approach of Phaethon. The weather during each observational run was mostly clear; however, on the nights of 11–13 November 2017 at Mt. Lemmon, USA, 24 November 2017 at Mt. Sobaek, Korea, 23, 27 November in Maidanak, Uzbekistan, and 15 December 2017 in Taiwan, we observed cirrus. To characterise the rotational status of Phaethon, time-series observations were performed, mostly using the Johnson R filter, because the combinations of the R band and the optical imagers provide the highest sensitivity for rocky bodies in the solar system. In addition, observing runs dedicated to the calibration of the datasets from different telescopes were performed using the LOAO 1.0 m telescope in January 2018. To calibrate all the data obtained from various telescopes, the same CCD fields during the observing run between November and the beginning of December in 2017 were taken on a single photometric system using the LOAO 1.0-m telescope.

\section{Data reduction and lightcurve analysis}
All data-reduction procedures were performed using the Image Reduction and Analysis Facility (IRAF) software package. Individual images were calibrated using standard processing routine of the IRAF task $noao.imred.ccdred.ccdproc$. Bias and dark frames with relatively large standard deviations were not used for our analysis. Twilight sky flats were acquired before sunrise and after sunset and combined to produce a master flat image for each night. The instrumental magnitudes of Phaethon were obtained using the IRAF $apphot$ package; the aperture radii were set to be equal to FWHM of the stellar profile on each frame in order to maximise the S/N ratio \citep{howell}. The lightcurve of Phaethon was constructed on the basis of the relative magnitude, which is the difference between the instrumental magnitude of the asteroid and the average magnitude of each comparison star. To choose a set of comparison stars, we used the dedicated photometric analysis software subsystem for asteroids, which is called the Asteroid Spin Analysis Package \citep[see][for more details]{kim}. This package helps to find appropriate comparison stars from single night images and to derive the spin parameters. In consequence, we selected three to five comparison stars with typical scatter of 0.01–0.02 magnitudes. The observation time (UT) was corrected for the light-travel time, and the influence of the distance from the Earth and the Sun was corrected.

   \begin{figure}
   \centering
   \includegraphics[width=\hsize]{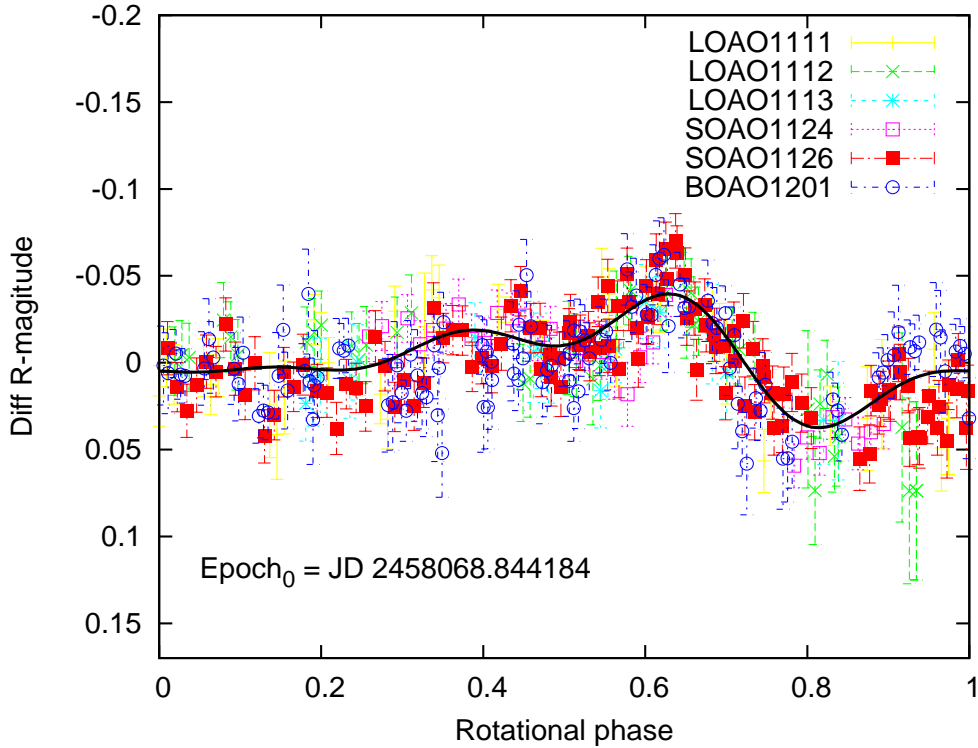}
      \caption{Composite lightcurve of Phaethon folded with the rotational period of 3.604 h at the zero epoch of JD 2458068.844184. The black solid line is a fit to the fourth-order Fourier model using the $F_\mathrm{\chi^2}$ method. Each data point represents observatories (see abbreviations in Table 1) and observing dates (MMDD).}
         \label{FigVibStab}
   \end{figure}

To determine the periodicity of the lightcurve, the fast chi-squared ($F_\mathrm{\chi^2}$) method \citep{palmer} was adopted. In addition, the result was confirmed via the discrete Fourier transform algorithm \citep{lenz}. These different techniques yield similar results, for example rotation periods of 3.6043 and 3.6039 h, respectively, which is consistent with previous lightcurve observations of Phaethon \citep{wisniewski,pravec,krugly,ansdell,warner2015,schmidt}.

The $F_\mathrm{\chi^2}$ technique presented here employs a Fourier series truncated at the harmonic $H$:
\begin{equation}
\Phi_H(\{A_{0...2H},f\},t) = A0 + \sum_{h=1...H}A_{2h-1}\sin(h2{\pi}ft) + A_{2h}\cos(h2{\pi}ft)
\end{equation}

We fit the fourth-order Fourier function with $A_\mathrm{0}$ = -0.00187142, $A_\mathrm{1}$ = -0.00899222, $A_\mathrm{2}$ = -0.0169267, $A_\mathrm{3}$ = -0.00845652, $A_\mathrm{4}$ = 0.00929936, $A_\mathrm{5}$ = 0.011695, $A_\mathrm{6}$ = -0.002285, $A_\mathrm{7}$ = -0.00194544, and $A_\mathrm{8}$ = -0.00699133. We also obtained the highest spectral power at P = 13.31869 cycles/day using the discrete Fourier transform algorithm. As a result, a rotational period of 3.6039 h was obtained, assuming a double-peaked lightcurve. We present the resultant composite lightcurve of Phaethon in Fig. 1, which folds with the period of 3.604 h at the epoch $t_\mathrm{0}$ of JD = 2458068.844184. We combined the data obtained from the SOAO, LOAO, and BOAO telescopes and computed the relative magnitudes according to the observations of comparison stars, using the observations conducted on 5–6 January 2018 as a reference for our calibration procedure.

The amplitude of the lightcurve computed via curve fitting (black solid line in Fig. 1) is $\Delta m = 0.075 \pm 0.035$. The peak-to-peak variations in magnitude are caused by the change in the apparent cross-section of the rotating tri-axial ellipsoid, with semi-axes a, b, and c, where a > b > c (rotating about the c axis). According to \citet{binzel1989}, the lightcurve amplitude varies as a function of the polar aspect viewing angle $\theta$ (the angle between the rotation axis and the line of sight): 
\begin{equation}
{\Delta}m = 2.5log(\frac{a}{b}) - 1.25log(\frac{a^2\cos^2{\theta}+c^2\sin^2{\theta}}{b^2\cos^2{\theta}+c^2\sin^2{\theta}})
\end{equation}

The lower limit of axis ratio a/b can be expressed as $a/b = 10^{0.4{\Delta}m}$, assuming an equatorial view ($\theta = 90\degr$). From this calculation, the lower bound for the a/b axis ratio of Phaethon is 1.07. 

\section{Shape model and pole orientation}
The lightcurve-inversion method \citep{kaasalainen2001a,kaasalainen2001b} is a powerful tool for acquiring the rotational status, including the spin orientation and the shape of asteroids, from the disk-integrated time-series photometric data. For this purpose, the lightcurve data obtained over three or four apparitions are essential. For this reason, we utilised as many lightcuves of Phaethon as possible, mainly based on our observations but also with data available in the literature from the Database of Asteroid Models from Inversion Techniques \citep{durech2010} and the Asteroid Lightcurve Database \citep{warner2009}. The detailed information and references of each lightcurve from the database are shown in Table A.1 in Appendix A. The total number of input datasets is 114 lightcurves, and the time span of the observations is 1994 to 2017. A first-period search using the $period\_scan$ programme was conducted between 1 and 24 h to find the global minimum ${\chi^2}$ value, and the results were scanned between 3.3 and 3.9 h with an interval coefficient of 0.8, which corresponds to $2.5\times 10^{-5}$ h to refine and find the unique sidereal period. The optimal solution was found at the sidereal period of P = 3.603957 h (see Fig. 2), which is consistent with a previous study \citep{hanus2016}. Once a unique solution for the sidereal rotational periods is determined, numerous shape models with the pole orientation are applied to find the pole pair ($\lambda, \beta$) by scanning the entire celestial sphere. Consequently, we found the lowest ${\chi^2}$ value near (308, -52) (see Fig. 3 and Table 3), which corresponds to the first pole orientation of \citet{hanus2016} preferred there due to a better fit to the thermal infrared data from Spitzer. There is a common practice to consider the solution as unique if there is only one pole solution that gives a significantly lower ${\chi^2}$ (by 10\%) than all others \citep{hanus2011}.

   \begin{figure}
   \centering
   \includegraphics[width=\hsize]{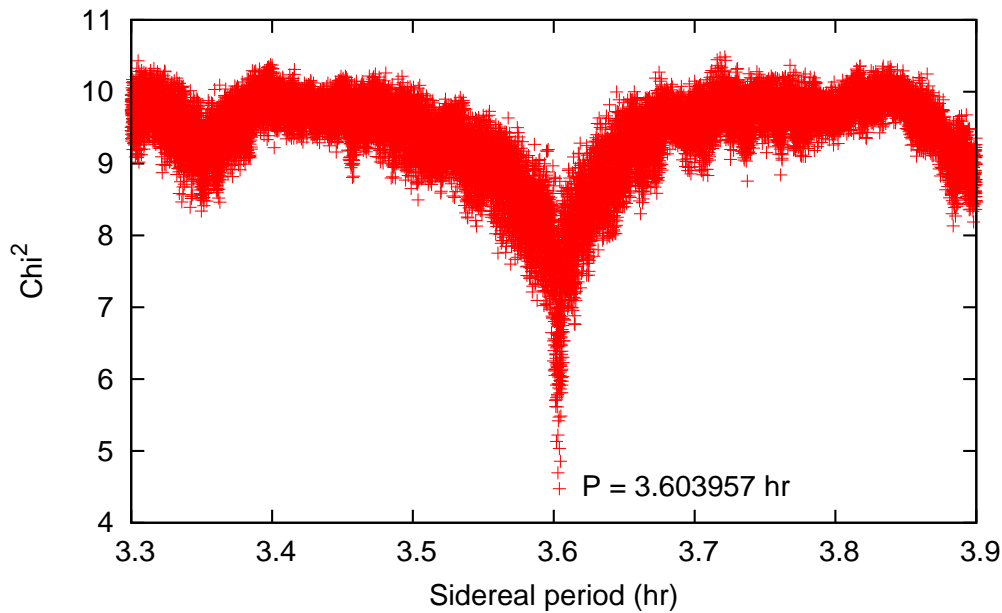}
      \caption{Periodograms of Phaethon obtained from the period scan programme. The rotation period subspace exhibits one prominent minimum corresponding to the period of P = 3.603957 h.}
         \label{FigVibStab}
   \end{figure}

   \begin{figure}
   \centering
   \includegraphics[width=\hsize, width=12cm, angle=270]{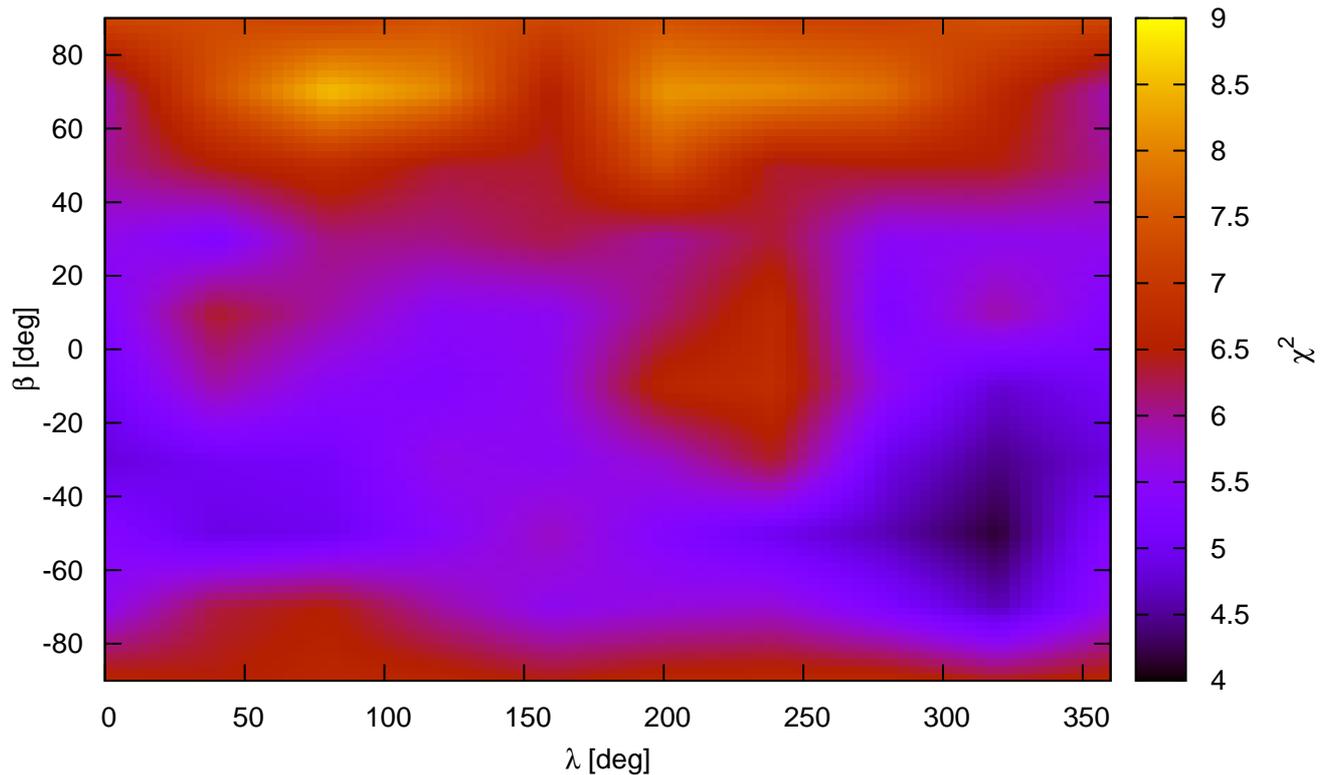}
      \caption{Pole solution distribution of Phaethon. The lowest ${\chi^2}$ value near (308, -52) was found from the lightcurve inversion method. $\lambda$ and $\beta$ are the ecliptic longitude and latitude of the pole orientation, respectively}
         \label{FigVibStab}
   \end{figure}

\begin{table*}
\caption{Sidereal rotational period and pole orientation of Phaethon.}
\label{table:3}
\centering
\begin{tabular}{c c c c c c}     
\hline\hline
$\lambda_\mathrm{1}$ & $\beta_\mathrm{1}$ & $\lambda_\mathrm{2}$ & $\beta_\mathrm{2}$ & $P_\mathrm{sid}$ & References  \\
(deg) & (deg) & (deg) & (deg) & (hr) &   \\
\hline
308 $\pm$ 10	& -52 $\pm$ 10 &  &  & 3.603957 & This work (LI) \\
322 $\pm$ 10 &	-40 $\pm$ 10 & & & 3.603956 & This work (SAGE) \\
319 $\pm$ 5 & -39 $\pm$ 5 & 84 $\pm$ 5 & -39 $\pm$ 5 & 3.603958 $\pm$ 0.000002 & \citet{hanus2016} \\
 & & 85 $\pm$ 13 & -20 $\pm$ 10 & 3.6032 $\pm$ 0.0008 & \citet{ansdell} \\
276 & -15 & 97 &-11 & 3.59060 & \citet{krugly} \\
\hline
\end{tabular}
\tablefoot{The ecliptic longitude ($\lambda$) and latitude ($\beta$) of the asteroid pole orientation, the sidereal rotational period ($P_\mathrm{sid}$), and references. Our solutions are derived from lightcurve inversion (LI) method and shaping asteroid models using the genetic evolution (SAGE) algorithm, respectively.}
\end{table*}

We present the 3D shape model of Phaethon based on the unique solution with a sidereal period of 3.603957 h and a pole orientation of (308, -52) (see Fig. 4). The actual value of the a/b ratio from the 3D model solution is 1.118. So we confirm the lower bound for the a/b axis ratio obtained from the lightcurve amplitude. In addition, the spin solution was confirmed by the model from the independent SAGE method \citep{bartczak}. The SAGE method is based on photometric data and uses a genetic evolution algorithm to fit the model’s shape and spin parameters to the lightcurves. Assuming homogeneous mass distribution, the spin axis of the resulting non-convex shape goes through the centre of mass and lies along the axis with the largest moment of inertia. The RMS (root mean square) values of the model fit from the lightcurve inversion (LI) method and SAGE algorithm are 0.02378 and 0.02738, respectively. Comparing two models from the composite lightcurves for a subset of the observation, the convex and non-convex models are generally similar to each other (see Fig. A.1). Although the SAGE model gives a better fit to explain the minima and maxima of the lightcurves in some data, for many other lightcurves the convex model fits much better than the SAGE model.

The non-convex shape model from the SAGE method is also shown in Fig. 4. The convex and non-convex models are practically not elongated in shape, which is a predictable result from the lightcurve amplitude. In comparison with the convex model, the non-convex one has some concavity features. In general, however, a non-convex model cannot be uniquely determined based on the photometric lightcurve only. Because it is possible to reconstruct the different shapes of concavities along the same line, these recesses produce a shadow effect due to one concavity of a different shape \citep{viikinkoski}. Recently, a number of works for the reconstruction of a non-convex model have been conducted giving weight to various sources (i.e. adaptive optics, lightcurve, and stellar occultation) \citep[see][and references therein]{hanus2017,viikinkoski}. There is one asteroid (3103 Eger) that has a non-convex model with photometry only \citep{durech2012}. Its non-convex shape model better fits the lightcurves that were observed at large phase angles than a convex one.

   \begin{figure}
   \centering
   \includegraphics[width=\hsize]{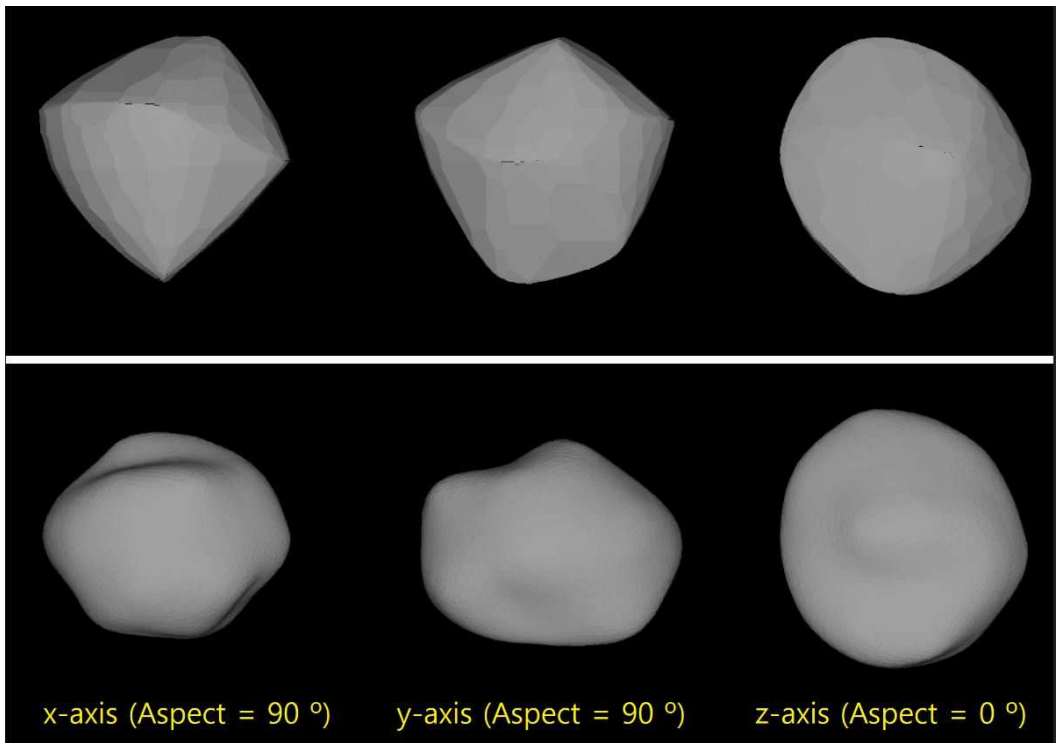}
      \caption{Three-dimensional shape model of Phaethon obtain from LI method (top) and the SAGE algorithm (bottom). The three views in both shape models correspond to the views from the positive x, y, z axes, respectively.}
         \label{FigVibStab}
   \end{figure}

\section{Conclusions}
The observation campaign for Phaethon was performed on the Asian and American continents owing to their favourable observation conditions at the end of 2017. We employed eight 1–2-m-class telescopes for a total of 22 nights between 11 November and 17 December 2017. The observation at the closest approach point on 16 December 2017 was conducted when the geometric configuration between the Earth and the asteroid with respect to the Sun changed dramatically. According to our observation datasets, we obtained the composite lightcurve of Phaethon, finding a synodic rotational period of $3.6039 \pm 0.0004$ h using two independent methods, and we calculated the lightcurve amplitude to be $0.095 \pm 0.035$, which is regarded as a nearly spherical shape. 

In addition, we derived sidereal rotational periods of 3.603957 and 3.603956 h and pole orientations of (308, -52) and (322, -40) in the ecliptic reference frame using the lightcurve-inversion method \citep{kaasalainen2001a,kaasalainen2001b} and the SAGE algorithm \citep{bartczak}, respectively. We also obtained the 3D shape model from both methods. According to \citet{taylor}, there is a concavity feature near the equator from Arecibo radar observations. However, the features are not obviously matched with our non-convex model. Apparently, the non-convex model contains many details that may lead to misinterpretation \citep{durech2012}. Because the RMS of the convex model is low (about 30\% in chi-square), we adopt the convex model as a pole solution for Phaethon. When we examine our pole solution in the geometric configuration, the aspect angle of Phaethon was 15 deg (almost pole-on view) to 40 deg during most of the 2017 apparition, then it quickly increased to 90 deg (edge-on view) during the close approaching phase. This could well explain the fact that the lightcurve amplitude of Phaethon during the 2017 apparition is relatively smaller than other lightcurve data obtained from previous apparitions. The new shape model and pole information that were obtained here are expected to be used not only for time-resolved spectroscopic and polarimetric observation of Phaethon but also to constrain the mission design for the DESTINY$^{+}$ science and engineering team, which is scheduled for 2022.

\begin{acknowledgements}
      MJK was supported by the Korea Astronomy and Space Science Institute. The work at Adam Mickiewicz University leading to these results has received funding from the European Union's Horizon 2020 Research and Innovation Programme, under Grant Agreement no 687378. ZYL was supported by grant number MOST 105-2112-M-008-002-MY3 and MOST 104-2112-M-259-006 from the Ministry of Science and Technology of Taiwan. AS, MK and IR were supported by the Targeted Financing Program BR05336383 Aerospace Committee of the Ministry of Defense and Aerospace Industry of the Republic of Kazakhstan.
\end{acknowledgements}

\bibliographystyle{aa} 
\bibliography{Phaethon_final_referee} 

\newpage
\begin{appendix}
\section{Additional table and figures}

\begin{longtable}{c c c c c c c c} 
    \caption{List of the lightcurve references from DAMIT and LCDB.} \\ \hline\hline
    \centering
    No & Epoch & N$_\mathrm{P}$ & $\alpha$ & r & $\Delta$ & Telescopes & References \\
    & [UT] & & [\degr] &  [AU]  & [AU] & & \\ \hline
    \endfirsthead
    \caption[]{List of the lightcurve references from DAMIT and LCDB (cont').} \\ \hline\hline
    \centering
    N & Epoch & N$_\mathrm{P}$ & $\alpha$ & r & $\Delta$ & Telescopes & References \\
    & [UT] & & [\degr] &  [AU]  & [AU] & & \\ \hline
    \endhead
    \hline
    \endfoot{}
 1 & 1994-11-02.1 & 22  & 25.5 & 1.82 & 1.04 & D65                  & \citet{hanus2016} \\                     
 2 & 1994-12-02.9 & 14  & 10.9 & 1.53 & 0.56 & D65                  & \citet{hanus2016} \\                     
 3 & 1994-12-04.1 & 17  & 11.2 & 1.51 & 0.54 & D65                  & \citet{hanus2016} \\                     
 4 & 1994-12-06.9 & 13  & 13.0 & 1.48 & 0.52 & D65                  & \citet{hanus2016} \\                     
 5 & 1994-12-27.3 & 76  & 48.0 & 1.22 & 0.44 & Lowell               & \citet{ansdell} \\                    
 6 & 1995-01-04.4 & 11  & 63.2 & 1.10 & 0.46 & UH88                 & \citet{ansdell} \\            
 7 & 1995-01-04.8 & 45  & 63.9 & 1.10 & 0.46 & D65                  & \citet{pravec} \\                     
 8 & 1995-01-05.4 & 79  & 64.9 & 1.09 & 0.46 & UH88                 & \citet{ansdell} \\            
 9 & 1997-11-01.1 & 88  & 48.3 & 1.32 & 0.78 & D65                  & \citet{pravec} \\                    
10 & 1997-11-02.1 & 80  & 48.9 & 1.31 & 0.76 & D65                  & \citet{pravec} \\                    
11 & 1997-11-11.6 & 39  & 57.0 & 1.18 & 0.56 & UH88                 & \citet{ansdell} \\             
12 & 1997-11-12.6 & 52  & 58.2 & 1.16 & 0.54 & UH88                 & \citet{ansdell} \\             
13 & 1997-11-21.6 & 48  & 74.1 & 1.02 & 0.39 & UH88                 & \citet{ansdell} \\             
14 & 1997-11-22.6 & 47  & 76.7 & 1.01 & 0.37 & UH88                 & \citet{ansdell} \\             
15 & 1997-11-25.6 & 24  & 85.7 & 0.95 & 0.34 & UH88                 & \citet{ansdell} \\             
16 & 1998-11-22.1 & 14  & 9.0  & 2.31 & 1.36 & IAC-80               & \citet{hanus2016} \\               
17 & 1998-11-23.1 & 16  & 9.2  & 2.31 & 1.36 & IAC-80               & \citet{hanus2016} \\               
18 & 1998-12-08.0 & 9   & 15.3 & 2.26 & 1.39 & IAC-80               & \citet{hanus2016} \\                
19 & 1998-12-09.0 & 15  & 15.8 & 2.25 & 1.40 & IAC-80               & \citet{hanus2016} \\               
20 & 2004-11-19.5 & 38  & 13.6 & 1.78 & 0.83 & UH88                 & \citet{ansdell} \\                   
21 & 2004-11-21.6 & 51  & 12.4 & 1.76 & 0.81 & UH88                 & \citet{ansdell} \\                   
22 & 2004-11-22.4 & 35  & 12.0 & 1.75 & 0.80 & UH88                 & \citet{ansdell} \\                   
23 & 2004-12-05.0 & 101 & 12.2 & 1.63 & 0.67 & D65                  & \citet{hanus2016} \\         
24 & 2004-12-05.3 & 41  & 12.4 & 1.63 & 0.67 & Badlands Observatory & \citet{hanus2016} \\
25 & 2004-12-11.0 & 148 & 18.1 & 1.57 & 0.64 & D65                  & \citet{hanus2016} \\         
26 & 2004-12-18.8 & 15  & 27.9 & 1.48 & 0.61 & D65                  & \citet{hanus2016} \\          
27 & 2007-11-17.2 & 47  & 44.6 & 1.28 & 0.51 & Modra                & \citet{hanus2016} \\                     
28 & 2007-11-28.2 & 96  & 54.1 & 1.13 & 0.29 & Modra                & \citet{hanus2016} \\                     
29 & 2007-12-04.1 & 232 & 69.9 & 1.03 & 0.18 & Modra                & \citet{hanus2016} \\           
30 & 2013-11-20.3 & 24  & 61.5 & 1.07 & 0.80 & UH88                 & \citet{ansdell} \\                   
31 & 2013-11-23.3 & 16  & 58.2 & 1.12 & 0.84 & UH88                 & \citet{ansdell} \\                  
32 & 2013-12-03.2 & 20  & 49.6 & 1.26 & 1.02 & Lowell               & \citet{ansdell} \\         
33 & 2013-12-11.3 & 36  & 44.6 & 1.37 & 1.18 & UH88                 & \citet{ansdell} \\                  
34 & 2014-11-27.3 & 89  & 9.3  & 1.82 & 0.85 & CS3-PDS              & \citet{warner2015} \\                  
35 & 2014-11-28.2 & 84  & 9.3  & 1.81 & 0.85 & CS3-PDS              & \citet{warner2015} \\                  
36 & 2014-11-28.4 & 58  & 9.3  & 1.81 & 0.84 & CS3-PDS              & \citet{warner2015} \\                  
37 & 2014-11-29.3 & 82  & 9.4  & 1.80 & 0.84 & CS3-PDS              & \citet{warner2015} \\                  
38 & 2014-11-29.5 & 27  & 9.5  & 1.80 & 0.84 & CS3-PDS              & \citet{warner2015} \\                 
39 & 2014-12-08.0 & 4   & 14.6 & 1.72 & 0.78 & CS3-PDS              & \citet{warner2015} \\                  
40 & 2014-12-10.1 & 91  & 16.4 & 1.71 & 0.78 & C2PU                 & \citet{hanus2016} \\                 
41 & 2014-12-11.9 & 92  & 18.1 & 1.69 & 0.77 & C2PU                 & \citet{hanus2016} \\                 
42 & 2014-12-14.2 & 52  & 20.4 & 1.67 & 0.77 & CS3-PDS              & \citet{warner2015} \\                  
43 & 2014-12-15.3 & 73  & 21.4 & 1.66 & 0.77 & CS3-PDS              & \citet{warner2015} \\                  
44 & 2015-01-13.9 & 54  & 48.0 & 1.32 & 0.83 & C2PU                 & \citet{hanus2016} \\    
45 & 2015-01-17.9 & 50  & 50.8 & 1.27 & 0.85 & C2PU                 & \citet{hanus2016} \\                 
46 & 2015-02-09.8 & 30  & 66.4 & 0.91 & 0.89 & C2PU                 & \citet{hanus2016} \\                 
47 & 2015-02-10.8 & 41  & 67.2 & 0.89 & 0.89 & C2PU                 & \citet{hanus2016} \\                 
48 & 2015-02-11.8 & 39  & 68.0 & 0.87 & 0.89 & C2PU                 & \citet{hanus2016} \\                 
49 & 2015-08-21.6 & 26  & 27.5 & 2.15 & 2.09 & UH88                 & \citet{hanus2016} \\                    
50 & 2015-09-08.6 & 22  & 26.6 & 2.24 & 1.90 & UH88                 & \citet{hanus2016} \\                    
51 & 2015-09-09.6 & 30  & 26.5 & 2.24 & 1.89 & UH88                 & \citet{hanus2016} \\                    
52 & 2015-10-08.5 & 21  & 20.0 & 2.33 & 1.60 & UH88                 & \citet{hanus2016} \\    
53 & 2016-11-02.2 & 49  & 33.9 & 1.49 & 0.69 & CS3-PDS              & \citet{warner2017} \\
54 & 2016-11-03.2 & 62  & 33.7 & 1.50 & 0.71 & CS3-PDS              & \citet{warner2017} \\ 
55 & 2016-11-04.2 & 119 & 33.5 & 1.51 & 0.72 & CS3-PDS              & \citet{warner2017} \\
56 & 2016-11-05.2 & 109 & 33.3 & 1.52 & 0.74 & CS3-PDS              & \citet{warner2017} \\
57 & 2017-11-26.3 & 89  & 29.0 & 1.31 & 0.39 & CS3-PDS              & \citet{warner2018} \\
58 & 2017-12-01.3 & 25  & 26.1 & 1.24 & 0.30 & CS3-PDS              & \citet{warner2018} \\
59 & 2017-12-01.4 & 24  & 26.0 & 1.24 & 0.29 & CS3-PDS              & \citet{warner2018} \\
60 & 2017-12-02.2 & 21  & 25.4 & 1.23 & 0.28 & CS3-PDS              & \citet{warner2018} \\ 
61 & 2017-12-02.3 & 12  & 25.4 & 1.23 & 0.28 & CS3-PDS              & \citet{warner2018} \\ 
62 & 2017-12-02.4 & 23  & 25.3 & 1.23 & 0.28 & CS3-PDS              & \citet{warner2018} \\
63 & 2017-12-17.0 & 424 & 66.8 & 1.01 & 0.06 & Burleith Observatory & \citet{schmidt} \\ 
\hline
\end{longtable}
\tablefoot{Epoch (UT date) corresponding to the mid-time of the observation, the solar phase angle ($\alpha$), the helicentric ($r$) and the topocentric distances ($\Delta$). Modification of Table A.1. from Hanu\u{s} et al. (2016) to include phase angle and recent observations.}

\begin{figure*}
   \centering
  \includegraphics[width=\textwidth, width=16cm]{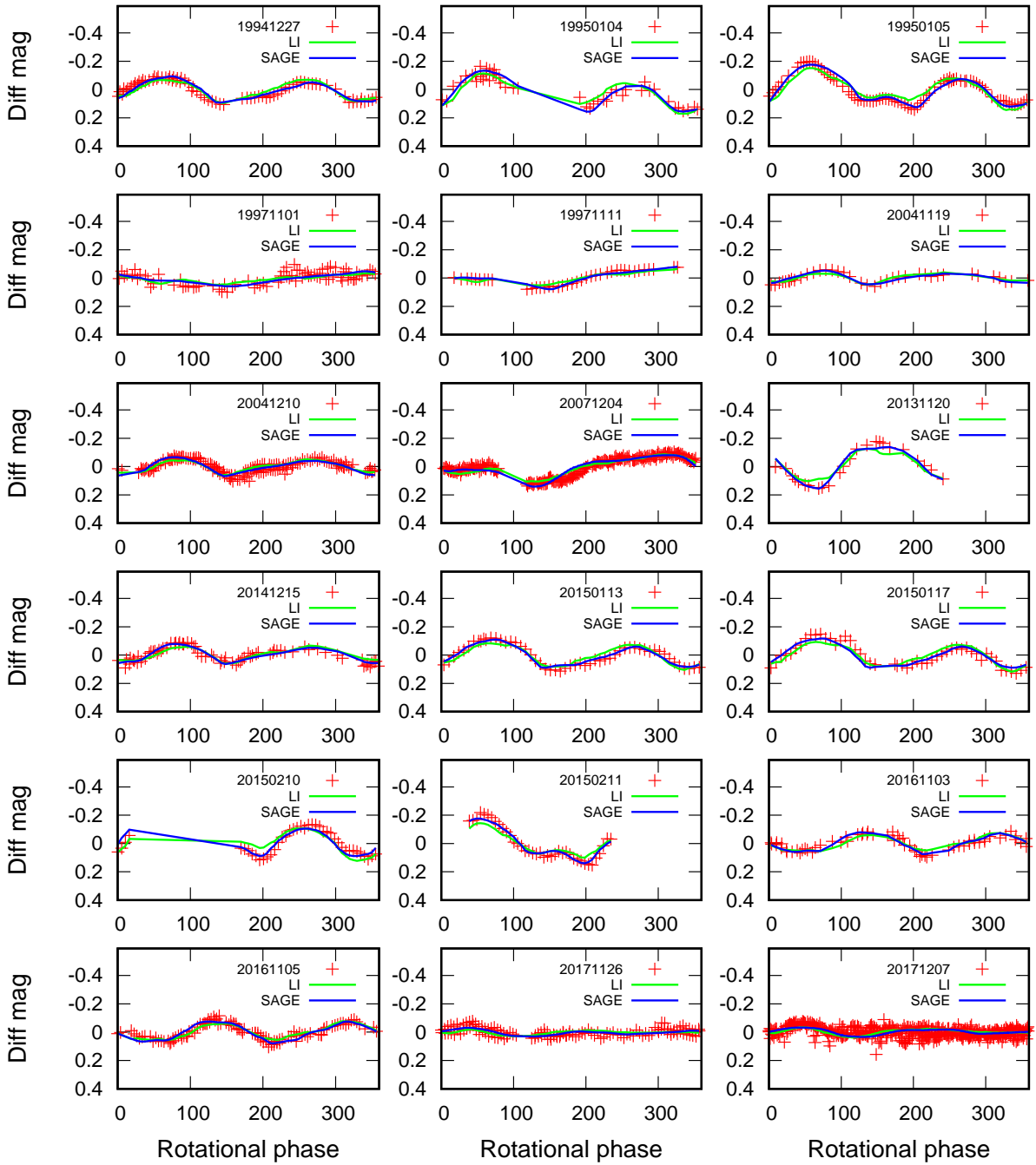}
  \caption{Comparison between the composite lightcurve (red cross) for a subset of the observation data (YYYYMMDD) and the model curves from the lightcurve inversion (LI) method (green fit) and SAGE algorithm (blue fit), respectively.}
\end{figure*}

\end{appendix}

\end{document}